# Neural network algorithm and its application in temperature control of distillation tower

Ningrui Zhao     Jinwei Lu

School of Chemical Engineering, East China University of Science and Technology, Shanghai 200237

# Abstract

Distillation process is a complex process of conduction, mass transfer and heat conduction, which is mainly manifested as follows: The mechanism is complex and changeable with uncertainty; the process is multivariate and strong coupling; the system is nonlinear, hysteresis and time-varying. Neural networks can perform effective learning based on corresponding samples, do not rely on fixed mechanisms, have the ability to approximate arbitrary nonlinear mappings, and can be used to establish system input and output models. The temperature system of the rectification tower has a complicated structure and high accuracy requirements. The neural network is used to control the temperature of the system, which satisfies the requirements of the production process. This article briefly describes the basic concepts and research progress of neural network and distillation tower temperature control, and systematically summarizes the application of neural network in distillation tower control, aiming to provide reference for the development of related industries.

Keywords：**Neural Networks, Control process, Distillation tower, Temperature parameter**

# 1 Introduction

Distillation is the most widely used mass transfer and heat transfer process in chemical production. Its model is complex and difficult to establish, and it is necessary to precisely control various parameters when it is desired to meet higher product purity. Among them, the composition and temperature of materials are critical. The detection of product composition has serious hysteresis, and the temperature measurement is convenient and fast, and its change is closely related to the product composition. Therefore, the temperature release of the rectification tower is a crucial parameter, but temperature control has coupling, hysteresis, and time-varying properties [1]. Traditional manual operation and conventional PID are difficult to achieve precise control.

Neural networks can perform effective learning based on corresponding samples, find corresponding laws like the human brain, do not rely on fixed mechanisms, and have the ability to approximate arbitrary nonlinear mappings [2], which can be used to establish input and output models of complex systems. Over the years, it has been widely used in many fields such as speech processing [3], image recognition [4], medical diagnosis [5], vehicle tracking [6], automatic control of chemical processes [7], renewable energy prediction [8], etc.

The application of neural network to the rectification system can well solve the difficult problems of the temperature control mechanism of the rectification tower such as complex and changeable and strong coupling. Many scholars have done a lot of research on the model establishment and control simulation of rectification tower and have achieved good results. This article will briefly describe the basic concepts and research progress of neural network and distillation tower temperature control, and systematically introduce the application of neural network in distillation tower control, such as PID neural network decoupling control, control based on BP neural network Method, control method based on RBF neural network, control method based on fuzzy neural network, and some optimization algorithm applications such as IGA (immune genetic algorithm) optimization, PSO (particle swarm algorithm) optimization and AFSA (artificial fish swarm algorithm) optimization .

## 2 Distillation tower and its temperature control characteristics

The bubble point of the solution is related to the pressure and composition. The total pressure and composition of each plate in the rectification tower are different. In the normal pressure or pressure distillation tower, the total pressure of each plate is not much different, and the temperature distribution of the whole tower Mainly depends on the composition of each board, the temperature gradually increases from top to bottom. In vacuum distillation, the pressure drop between the top and the bottom of the tower is not negligible compared with the absolute pressure at the top of the tower, and it may be several times larger, which will affect the temperature of the entire tower.

The distillation process is a complex mass transfer and heat transfer process, and its temperature control characteristics are as follows: the mechanism is complex and changeable with uncertainty; the process is multivariate and strongly coupled [9]; the system has nonlinearity, hysteresis and time-varying properties [1]. Therefore, the use of neural network to control the system temperature can better meet the requirements of the production process.

The changes and connections of various parameters in the distillation column are based on two criteria: material balance and energy balance. The main disturbance factors of the rectification tower temperature are: the amount and composition of the feed, the amount and composition of the top and bottom sections; the enthalpy and temperature of the feed, the heat load of the reboiler and the condenser, the reflux ratio, and the ambient temperature [10], etc.

When the purity of the top product is very high, the temperature change in the top section of the tower is very small, so when the top temperature has a noticeable change, the fluctuation of the distillate composition has already exceeded the allowable range, so the top temperature cannot be used to control the distillate Composition, the temperature sensing element can be placed on the sensitive board to detect the interference of the distillation operation earlier [11]. Therefore, the neuron input layer variable in the neural network can be the temperature of the sensitive plate of the distillation tower, and the output variable can be the reboiler heating steam temperature control valve, so as to control the temperature of the sensitive plate [2]. Similarly, if the two ends are controlled, the reflux at the top of the tower is used to control the temperature at the top of the tower, and the heating at the bottom of the tower is used to control the temperature at the bottom of the tower to achieve the purpose of controlling product quality [12]. The temperature control diagram of the distillation tower is shown in Figure 1.

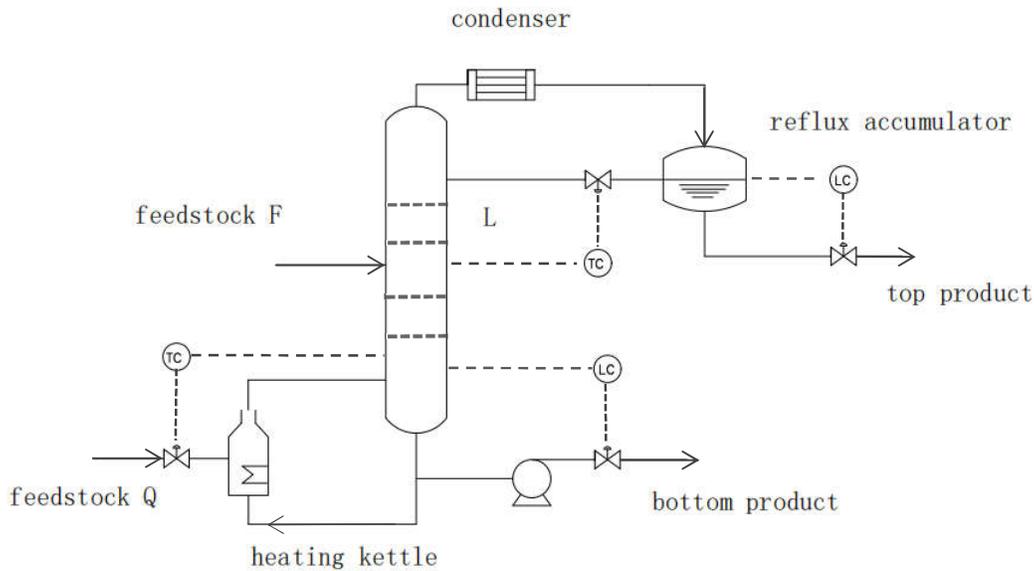

**Figure 1 Temperature control diagram of distillation tower**

## 3 Neural network and its application

A neural network is composed of multiple neurons. A neuron is a non-linear structure. Each neuron has input and output. A superimposed neural network is composed of multiple neurons, generally including input layer X and hidden units. And output layer Y [13]. The basic types of neural networks include single-layer forward networks, multilayer forward networks such as BP (Back Propagation) neural network, RBF (Radial Basis Function) neural network, feedback networks such as Hopfield Neural Network and other types [14].

The learning process of the neural network is carried out in the form of error back propagation, and the error back propagation algorithm is based on gradient descent. Assuming that the training results are correct, the weights will not change, otherwise, certain adjustments will be made. The multi-layer neural network propagates errors in the reverse direction from the output layer to the input layer for weight adjustment. The training process of the neural network is a process of parameter optimization. Based on gradient descent, it may be a local optimal solution but not necessarily the global optimal solution. Therefore, in the actual parameter tuning process, it often requires multiple trials to get a better solution. To deal with the non-convex non-linear structure of the neural network's high-dimensional parameter space, researchers have proposed distributional robust optimization algorithms [15] to search for the optimal solution.

Theoretically, deep neural networks can approximate any linear or non-linear mapping functions by simulating the interactive reaction of biological nervous system to real world things. It can be used to establish the input and output model of static systems, as well as to model the interactive behavior of intelligent agents in dynamic systems [16]. The neural network can learn effectively according to the corresponding samples, find the corresponding laws like the human brain, does not rely on a fixed mechanism, only needs to give enough samples for its learning, and has strong fault tolerance and adaptive learning capabilities.

# 4 Application of Neural Network in Temperature Control of Distillation Tower

## 4.1 PID neural network decoupling control

For a rectifying tower, a multi-variable coupling system with complex dynamic characteristics, although the traditional decoupling control can achieve decoupling control of a specific production process, the drawbacks of relying too much on accurate mathematical models make it control when the operating point changes. The effect is not satisfactory [17]. WANG [1] proposed to use neural network PID (Proportional-Integral-Derivative Neural Network, PIDNN) algorithm to design distillation column temperature controller by using the advantages of neural network's strong adaptive ability and fault tolerance ability.

Through the simulation of the traditional PID in the rectification tower temperature control system, it can be seen that when the system is disturbed, the system can finally reach the stable output of the control target, but the response of the top and bottom of the tower is lagging, and when the top of the tower is disturbed, the temperature of the bottom of the tower is also Oscillation occurs. After the bottom of the tower is disturbed, the temperature at the top of the tower also oscillates. Although the disturbance is quickly suppressed, the frequent oscillations caused by the coupling relationship in the actual device are unacceptable in production [18].

In the temperature control process with pure hysteresis characteristics, conventional PID control is difficult to meet the requirements of control accuracy, and process control is difficult to describe with mathematical models, and neural networks do not need to clarify the system mechanism, and the laws can be mastered according to learning and training.

The neuron network PID is a multi-layer forward neuron network that combines PID control law and neural network. According to the PID control structure, the network is divided into three layers, namely the input layer, the hidden layer and the output layer. Among them, PID's proportion, integral and differential are contained in the hidden layer as neurons. For a multivariate system with m inputs and n outputs, the network structure can be expressed as 2m×3m×2n, and the network layers are connected by network weights, and they are cross-connected from the hidden layer to the output layer. , Which constitutes the basic form of the decoupling controller of the multivariate system. Through the self-learning and adjustment of the network, it continuously adapts to the needs of the objective function to achieve adaptive decoupling control [19].

Figure 2 shows a binary variable neuron network decoupling control system, where R1 and R2 are target values, U1 and U2 are the control variables, and Y1 and Y2 are the current actual values. The neural network PID adjusts the inertia weight value through the back-propagation learning algorithm, and achieves the purpose of generalized decoupling through the comprehensive action of three neurons [1]. The disadvantage is that the selection of the initial value of the network weight depends on a large amount of actual adjustment experience, and the multi-variable complex control system similar to the distillation tower has many uncertain factors, and the neural network itself has a slow learning rate and poor dynamic performance. Such shortcomings cannot eliminate the static coupling of the controlled object, and it is not easy to obtain the ideal dynamic control effect. Therefore, most existing researches use improved particle swarm optimization (PSO) to combine with it.

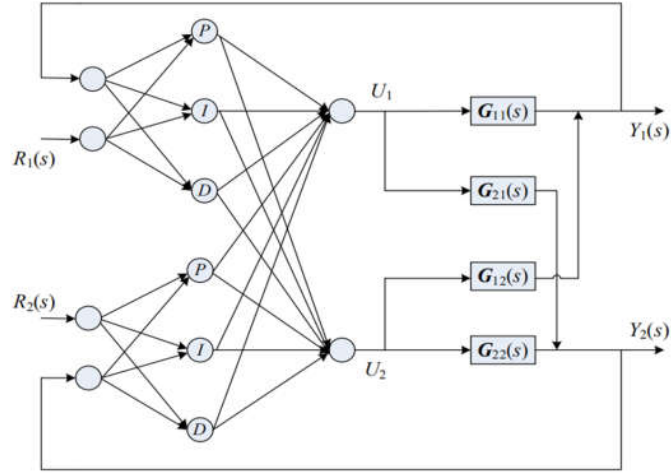

**Figure 2 Neural network decoupling control system**

## 4.2 BP neural network

BP (back propagation) neural network is a multi-layer feedforward neural network trained according to the error back propagation algorithm, and is currently the earliest and most widely used neural network [20]. It uses gradient search technology to minimize the error mean square error between the actual output value of the network and the expected output value. When the BP neural network is used in the entire distillation temperature control process, the neuron input layer variable can be set to the temperature of the sensitive plate of the distillation tower, and the output variable is the temperature of the reboiler heating medium, and the re-boiling is adjusted through the control valve The flow rate of the heating steam of the heater, so as to realize the control of the temperature of the sensitive plate.

Its outstanding advantage is that it has a strong nonlinear mapping ability and a flexible network structure. The number of intermediate layers of the network and the number of neurons in each layer can be set arbitrarily according to the specific situation, and its performance varies with the difference of the structure [20].

Liu Jian et al. proposed a temperature control system based on BP neural network, which provided an effective control scheme for the rectification tower that lacks precise mathematical models and nonlinearity, using four layers of l×25×20×1BP Neural Networks [21]. In order to control the composition of the bottom and top products of the tower, a dynamic mechanism model of the tower, that is, the forward model of the neural network, is established through the relationship between the heat and mass balance in the tower. The forward model uses a multi-layer feedforward neural network to express the positive dynamics of the system through training or learning. When there is a deviation between the actual output and the expected output, the error is propagated back. The neuron input layer variable is the temperature of the sensitive plate of the distillation tower, and the output variable is the temperature of the reboiler heating steam. The reboiler is adjusted by the control valve. The flow of heating steam can control the temperature of the sensitive plate.

But the BP neural network also has the following major defects:

① The learning speed is slow. Even for a simple problem, it usually takes hundreds or even thousands of learning to converge.

② It is easy to fall into a local minimum.

③ There is no corresponding theoretical guidance for the selection of the number of network layers and the number of neurons.

④ The network promotion ability is limited.

Therefore, many improvement studies have emerged, such as fuzzy BP neural network, artificial fish school algorithm (AFSA) optimized BP neural network and so on.

## 4.3 RBF (Radial Basis Function) Neural Network

The RBF radial basis function neural network structure is three-layer forward, each layer represents a specific function, the input layer has perceptual ability, the nodes of the hidden layer and the final output layer have computational performance, and the hidden layer passes Using nonlinear optimization strategy to adjust the parameters of its node radial basis function, the learning speed is slow, while the output layer adjusts the output weight through the linear optimization strategy, and the learning speed is fast [22]. It contains a non-linear mapping from input to output layer, and the hidden layer is a linear mapping to output. This feature makes its learning speed very fast, and it can also solve the local minimum.

The learning of RBF neural network is divided into forward learning algorithm and backward learning algorithm. The former is mainly used to calculate the control output, and the latter is mainly used to modify the network connection weights [23].

Compared with BP neural network in the application process, it shows many advantages:

① The RBF network can fit random continuous functions within any required accuracy range;

② RBF network has relatively fast learning efficiency;

③The RBF network has a good ability to solve the shortcomings contained in the local minimum.

ZHANG [24] carried out the RBF neural network identification simulation study of the top temperature of the rectification tower, and changed the size of the return flow to obtain the transfer function of the corresponding change in the top temperature. When a specific value of the system is reached, the system output is performed Identify. The results show that the actual output and the RBF neural network identification output are very small and the error is small. It is verified that the RBF neural identification method can be used to identify the system to obtain good tracking results, and the real-time output value of the object can be obtained, so that the system can complete Online control.

At the same time, he also carried out the RBF-PID control simulation study of the top temperature of the tower, and also selected the transfer function of the transfer function of the temperature change at the top of the rectification tower caused by the change of the reflux rate as the research object, and used the method of connecting the RBF and the PID controller. Control the temperature at the top of the distillation tower [24]. The results show that the overshoot and the adjustment time are much smaller than those of PID control, which can achieve good control of the tower top temperature and make it run safely and stably, thereby ensuring the quality and output of the tower top products.

The conventional RBF neural network also has its shortcomings. It has a fixed network structure and the number of hidden layer nodes must be determined in advance. The learning of the network is achieved by changing the central basis vector of the Gaussian function of the hidden layer nodes and the weights from the hidden layer to the output layer. Due to the uncertainty of the MIMO system model and the time-varying parameters, it is difficult for the RBF neural network with a fixed structure to achieve a global recognition effect [25]. TENG W F proposed an algorithm that combines dynamic RBF neural network and single neuron PID to decouple the distillation tower system. The nearest neighbor algorithm is used to construct the optimal dynamic RBF neural network to achieve the desired When the target is the target, the hidden layer has the least number of nodes. The dynamic RBF neural network is used to identify the system information. The obtained Jacobian information is used for single neuron PID online auto-tuning. A satisfactory decoupling effect is obtained and the system speed is better solved. The balance between accuracy and accuracy [26].

## 4.4 Fuzzy Neural Network

Neural networks and fuzzy systems are both nonlinear dynamic systems, which are model-free controllers, and are commonly used to deal with uncertain, nonlinear and other uncertain problems. But both have their own advantages and disadvantages: Fuzzy systems use anthropomorphic thinking patterns to extract knowledge and simple reasoning, but lack self-learning and self-adaptive capabilities; neural networks can effectively learn based on corresponding samples and can realize parallel computing Distributed storage of information and information, while having strong fault tolerance and adaptive learning capabilities, but it cannot make good use of the existing experience knowledge (generally only the initial value is zero or a random value), which may lead to a very long network training time Long, the network training falls into an non-required local extremum [27].

Fuzzy neural network is to combine the fuzzy system with the neural network and convert the fuzzy system into the corresponding neural network, thereby improving the expression ability and learning ability of the system [28].

GU et al. [14] proposed a fuzzy neural network controller, which takes the temperature, flow, and liquid level of the distillation tower as input, and the opening of the heat conduction oil valve as output. Through adaptive learning and fuzzy processing of manual operation, it can achieve Intelligent control of rectification tower temperature. The first step is to calculate and determine the parameters and formulas of the five-layer fuzzy neural network. The structure of the fuzzy neural network is shown in Figure 3. The first layer is the input layer, which buffers the input signal. The number of nodes is 3, and the second layer is the fuzzification layer. Each node represents a variable value. Its function is to calculate the membership degree of each input component belonging to the fuzzy set of each variable value Function, the third layer is the fuzzy rule layer. Each node represents a fuzzy rule. Its function is to match the antecedent of the fuzzy rule and calculate the applicability of each rule. The fourth layer is to realize normalized calculation, and the fifth layer is the output layer to realize clear calculation. The second step gives the learning algorithm for parameter adjustment. Since the fuzzy division number of the input component has been determined, the only parameters that need to be learned are the connection weight of the last layer, and the central value and width of the membership function of the second layer. According to the number of input nodes, the number of fuzzy divisions, the type of membership function and the processed sample data, the neural network classifier can calculate the fuzzy membership function and fuzzy rules through the second step of the above learning algorithm to complete the training . After the training is completed, given the input, the classifier can calculate the output value and select the output category through the output function. Training the generated FIS through a neural network, it is found that when the number of training is 6, the error is close to 0, and the result is fast convergence and small error. The neural network structure and various parameters are more suitable. The results show that the scheme can imitate manual operation, and the accuracy of intelligent learning is very high[14].

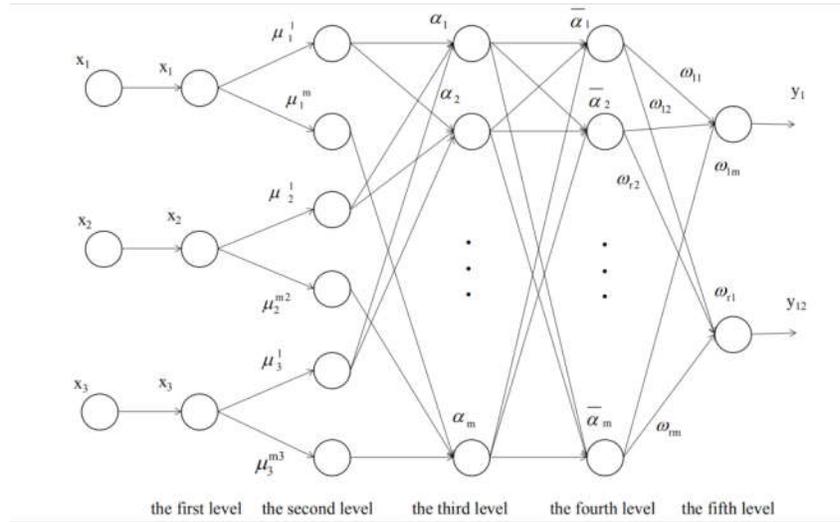

**Figure 3 Fuzzy Neural Network**

WANG et al. [29] used a fuzzy RBF neural network controller on the basis of the cascade control system to improve the control effect of the tower kettle temperature. The structure diagram of the fuzzy neural network controller is shown in Figure 4. The fuzzy neural network has a two-input three-output structure: the input is the temperature error and the temperature error change rate; the output is the three parameters Kp, Ki and Kd of the PID controller. The fuzzy logic is designed as follows: before fuzzy inference, the precise input quantity needs to be converted into fuzzy quantity, and then the fuzzy control rules of Kp, Ki, Kd are designed according to the experience of the operator. The next step is fuzzy inference and defuzzification. Fuzzy inference is According to the established fuzzy rules and input variables, the fuzzy result is inferred. The result of the inference is the fuzzy amount, which needs to be converted into a clear value by defuzzification. The defuzzification adopts the most common weighted average method, and then the linear transformation method is used. Finally, the actual control value, through the construction of the RBF fuzzy neural network and the design of the learning algorithm and MATLAB simulation, it is concluded that the control effect of this set of control schemes is fast, the overshoot is small, the adjustment time is short, the steady state error is small, and the comprehensive control level is high.

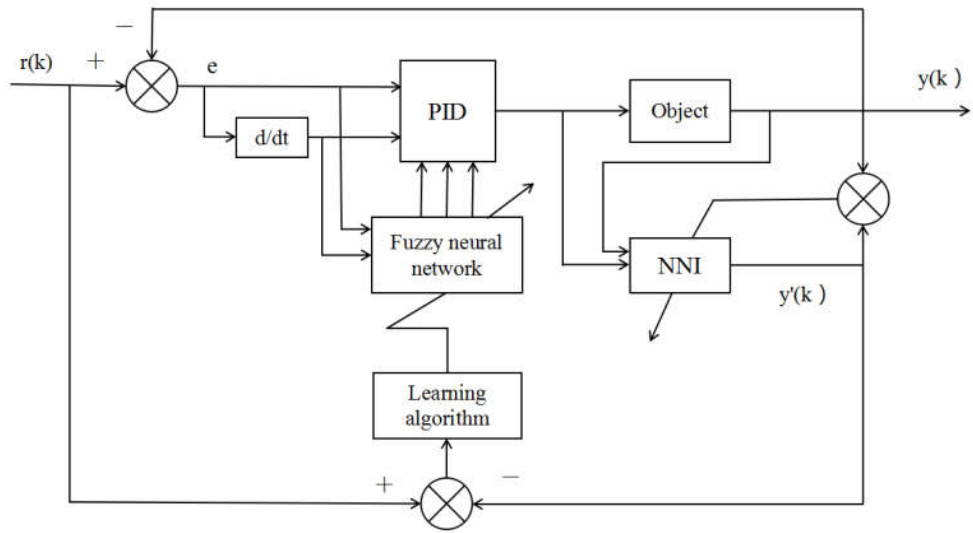

**Figure 4 Fuzzy neural network controller**

## 4.5 Optimization algorithm

### 4.5.1 IGA (Immune Genetic Optimization Algorithm)

Based on the existing theories of "natural selection" and "biological evolution" in the field of biology, Holland first proposed the "genetic algorithm" theory with significant influence. This algorithm has the characteristics of implicit parallelism, global optimization, and self-adaptation, so it has been used as an effective tool for nonlinear optimization problems, and it has been implemented in the distillation tower optimization system. At the same time, it participates in neural network learning behaviors, so Problems such as multi-objective optimization and non-polynomial completeness have been successfully handled, and satisfactory results have been obtained [30].

However, the pure genetic algorithm has a poor effect on discrete optimization problems, and it is easy to fall into local optimization, and particle swarm optimization (PSO) has more advantages in this respect [31].

This shortcoming of genetic algorithm can be improved in conjunction with immune operations. In the immune system, there is a feedback mechanism that can perform two different tasks at the same time; one is the emergence of external substances in response, and the other is to quickly stabilize the immune system. The diversity of antibodies can improve the global search capability of the genetic algorithm (IGA) and ensure that it does not fall into the local optimal solution, and the self-regulating mechanism can improve the local search capability of the genetic algorithm (IGA) [32].

GUO X Q et al. proposed a fuzzy controller based on immune genetic algorithm optimization, which uses the global search function of immune genetic algorithm and the self-learning ability of neurons to improve the control accuracy and anti-interference ability of the fuzzy controller. IGA algorithm steps: ①Create the initial population A1; ②Determine whether it meets the conditions, and stop if it meets the conditions; ③Perform crossover and mutation operations on the current population to obtain the next generation; ④Perform antibody and immune operations, and jump to ② [33]. The controller is used in the simulation of a full-stage distillation tower model. The simulation results show that the distillation tower is a very slow self-balancing system, which produces a population of 20 chromosomes and can obtain various steady-state data of the distillation tower. The controller can effectively eliminate static errors, and also has good robustness during the control transition process.

### 4.5.2 Particle Swarm Optimization Algorithm (PSO)

For the shortcomings of backpropagation algorithms and other intelligent optimization algorithms that converge slowly and easily fall into local extremes, particle swarm algorithms have been more widely used. The algorithm achieves global optimization by simulating the group behavior of birds looking for food, and other evolutionary algorithms In comparison, it has the advantages of simple and easy-to-understand concepts, easy program implementation, and fewer parameters [34]. The PSO algorithm parameters mainly include inertia weight, learning factor, maximum speed, and fitness function. The algorithm initializes a group of random particles, and determines the optimal position of the entire population of particles and the optimal position of individual particles according to the optimization goal. In the continuous iteration process, the optimal solution generated by the particle search is called the individual extreme point. The optimal solution is called the global extreme point [35]. Each particle continuously optimizes and updates its position by tracking two extreme points, and tends to the global extreme point until it reaches the specified number of iterations or the target value. As the number of iterations increases, the velocity of each particle approaches zero. At this time, the algorithm will fall into a local extreme. In order to avoid premature convergence of the algorithm, many scholars have proposed many improvements, including adding inertial weights and convergence factors, and referring to genetics. Evolutionary algorithm, improved topological structure, construction of new organizational structure, review degree update formula based on multi-objective optimization distillation system, and combined with various advanced algorithms to produce new particle swarm algorithm [35], such as WANG [1] proposed Gauss Improved PSO algorithm PID neural network decoupling control for white noise interference mutation, ASTROM et al. [16] proposed neural network PID decoupling control for chaotic particle swarm optimization, CHEN et al. [17] proposed PSO optimization neural network with speed reset strategy Network, and KENNEDY[18] proposed that periodic mutation operations can be introduced to avoid premature convergence of the algorithm.

### 4.5.3 AFSA (Artificial Fish School Optimization Algorithm)

AFSA (Artificial Fish School Optimization Algorithm) is a simulation algorithm derived from the life habits of fish schools. Its basic idea is: fish schools will always move toward areas with high food concentrations in the waters, thereby performing foraging behavior. As a result, artificial fish clusters are constructed for global optimization, which can effectively solve the situation that the BP neural network backpropagation commonly used steepest descent method is easy to fall into the maximum value of the area [36].

WANG et al. [37] proposed the BP algorithm fuzzy PID controller based on AFSA optimization to overcome the shortcomings of the steepest descent method in the back propagation process, which is easy to fall into the local optimal solution, and can obtain the optimal solution of PID parameters. Its AFSA algorithm outputs The reciprocal of the mean square error is used as the fitness function (food concentration) of the artificial fish swarm algorithm to determine the foraging behavior, grouping behavior, tail-catch behavior algorithm and parameters [37]. The simulation results show that the number of iterations of the AFSA-BP algorithm is small and can be close to the given value. When the relevant PID parameters are applied in the DMF recovery system, the temperature of the distillation column can be kept stable.

## 5 Summary and Outlook

Because the temperature control of the distillation tower is difficult to model, and the parameters have the characteristics of non-linearity, coupling, and time-varying, there are so many neural network methods to control the temperature of the distillation tower, which can accurately and effectively control the two ends of the distillation tower. Temperature, solve the problems of control lag and difficulty in stability. Among them, BP neural network is widely used in the temperature control of distillation towers, but with the continuous advancement of technology, people have found that it has problems such as slow learning speed, slow convergence, and easy to fall into local minimums. In addition, the number of samples is limited, while RBF neural network solves these defects of BP neural network well. The emergence of fuzzy neural network application research has improved the problem of long training time of neural networks, and simulated human thinking to extract knowledge. There are also many PID decoupling control strategies using neural networks to solve the problems of temperature coupling at the top of the tower and coupling of temperature with other parameters. The combination of optimization algorithms such as IGA (Immune Genetic Algorithm), PSO (Particle Swarm Algorithm) and AFSA (Artificial Fish Swarm Algorithm) and neural network makes temperature control more perfect. Among them, improved particle swarm algorithm is currently used in distillation tower temperature control More recognition in the industry. The application of these studies in large-scale production can greatly improve product quality indicators and produce greater economic benefits. However, most of the research done now relies on the results obtained by simulation in MATLAB. The effect of putting into operation in the actual factory is still unclear. I hope that the application of neural network in controlling the temperature of the rectification column can be deeply studied in the future, and applied in practice.

# References

[1] Wang HS. Modeling of Distillation Column and Adaptive Decoupling Control of Temperature [D]. Yanshan University, 2017. (in Chinese)

[2] Willis MJ, Montague GA, Di Massimo C, Morris A J and Tham MT. "Artificial neural networks and their application in process engineering," IEE Colloquium on Neural Networks for Systems: Principles and Applications, London, UK, 1991, pp. 7/1-7/4.

[3] Gao J, Chakraborty D, Tembine H, Olaleye O. "Nonparallel Emotional Speech Conversion," INTERSPEECH 2019, Graz, Austria, September 2019.

[4] Wu S, Rupprecht C, Vedaldi A. "Unsupervised Learning of Probably Symmetric Deformable 3D Objects From Images in the Wild," 2020 IEEE/CVF Conference on Computer Vision and Pattern Recognition (CVPR), Seattle, WA, USA, 2020, pp. 1-10, doi: 10.1109/CVPR42600.2020.00008.

[5] Aljurayfani M, Alghernas S, Shargabi A. "Medical Self-Diagnostic System Using Artificial Neural Networks," 2019 International Conference on Computer and Information Sciences (ICCIS), Sakaka, Saudi Arabia, 2019, pp. 1-5, doi: 10.1109/ICCISci.2019.8716386.

[6] Gao J, Tembine H. "Distributed mean-field-type filter for vehicle tracking," 2017 American Control Conference (ACC), Seattle, WA, 2017, pp. 4454-4459, doi: 10.23919/ACC.2017.7963641.

[7] Wu H, Zhao J. "Deep convolutional neural network model based chemical process fault diagnosis." Comput. Chem. Eng. 115 (2018): 185-197.

[8] Gao J, Chongfuangprinya P, Ye Y, Yang B. "A Three-Layer Hybrid Model for Wind Power Prediction," 2020 IEEE Power & Energy Society General Meeting (PESGM), Montreal, QC, 2020, pp. 1-5, doi: 10.1109/PESGM41954.2020.9281489.

[9] Du L, Cao JT, Li SC. Temperature Control of Distillation Tower Based on Improved Dynamic Neural Network[J]. Industry Control and Applications, 36(1): 25-31. (in Chinese)

[10] Huang JQ, Lewis FL. Neural-network Predictive Control for Nonlinear

Dynamics with Time-delay. IEEE Trans on neural networks, 2003, 2(14):377-389.

[11] Xu Y, Chuang KT. Design of a Process for Production of Isopropyl Alcohol by Hydration of Propylene in a Catalytic Distillation Column. Chemical Engineering Research and Design, 2002, 80(6): 686-694.

[12] Safavi AA, Romagnoli JA. Application of Wavelet-based Neural Networks to the Modeling and Optimization of an Experimental Distillation Column. Engineering Applications of Artificial Intelligence, 1997.

[13] Hagan MT, Demuth HB, Beale MH. Neural Network Design. Beijing: Chemical Industry Press. 2002.

[14] Gu YK, Yang CG, Wang HQ, Wang JB. Design of Distillation Column Temperature Controller Based on Fuzzy Neural Network Classifier[J]. Journal of Hefei University of Technology, 2012, 35(1): 41-44. (in Chinese)

[15] Gao J, Xu Y, Barreiro-Gomez J, Ndong M, Smyrnakis M, Tembine H. (September 5th, 2018) Distributionally Robust Optimization. In Jan Valdman, Optimization Algorithms, IntechOpen. DOI: 10.5772/intechopen.76686. ISBN: 978-1-78923-677-4

[16] Bauso D, Gao J, Tembine H. Distributionally Robust Games: f-Divergence and Learning, 11th EAI International Conference on Performance Evaluation Methodologies and Tools (VALUETOOLS), Venice, Italy, Dec 2017

[17] Takao K, Yamamoto T. A Design of Model Driven cascade PID Controllers Using a Neural Network. Conference on Neural Networks, 2003, 14(3): 22-27.

[18] Astrom KJ, Hagglund T. Automatic Tuning of PID Controllers. Research Triangle Park, North Carolina: Instrument Society of America, 1988.

[19] Chen SY, Lin FG. Decentralized PID Neural Network Control for Five Degree of Freedom Active Magnetic bearing, Engineering Applications of Artificial Intelligence, 2012, 26(7): 962-973.

[20] Kennedy J. Particle swarm optimization[C]. Proceedings of IEEE International conference on Neural Networks, 1995: 1942-1948.

[21] Liu J, Wang QL. Application of Neural Network in the Temperature Control System of Rectifying Tower[J]. Science and Technology Consulting Herald,

2007. (in Chinese)

[22] Wang HB, Yang XL, Wang HY. An improved learning algorithm for RBF neural networks. Systems Engineering and Electronics, 2002, 24(6): 103-105.

[23] Yang L, Ren XM, Huang H. Application of Self Tuning PID Controller Based on RBF Network. Computer Simulation，2006，23(1): 270-273.

[24] Zhang Y. Application Research of Neural Network in Distillation Column Control System[D]. Hebei University of Science and Technology, 2014. (in Chinese)

[25] Guillen A, Pomares H, Rojas I, Prieto A, Gonzalez J, Valenzuela O, Herrera LJ. Using Fuzzy Logic to Improve a Clustering Technique for Function Approximation. A new clustering technique for function approximation[J]. Neural Networks, 2002, 13(1): 132-142.

[26] Teng WF. Study on Decoupling Control Based on PSO Neural Network and Its Application in The Distillation[D]. Zhejiang Sci-Tech University, 2015. (in Chinese)

[27] Labiod S, Guerra TM. Adaptive Fuzzy Control of a Class of SISO Nonaffine Nonlinear Systems. Fuzzy Sets and Systems. 2007, 158: 1126-1137.

[28] Pehlivanoglu YV. A New Particle Swarm Optimization Method Enhanced with a Periodic Mutation Strategy and Neural Networks[J]. IEEE Transaction on Evolutionary Computation, 2013, 17(3): 436-452.

[29] Wang HQ, Cheng ZC. Design and Application of Fuzzy Neural Network Controller in DMF Recovery System[J]. Chemical automation and instrumentation, 2015, 42(3): 292-309. (in Chinese)

[30] Zhang S, Gao W, Qi M, Yu WH, Wang HH. A Review of Optimization Rectification Systems Based on Multi-objective[J]. Chemical Industry and Engineering Progress, 38(1):1-9. (in Chinese)

[31] Ourique CO, Biscaia EC, Pinto JC. The Use of Particle Swarm Optimization for Dynamical Analysis in Chemical Processes[J]. Computers & Chemical Engineering, 2002, 26(12):1783-1793.

[32] Jiao LC, Wang L. A novel genetic algorithm based on immunity[J]. IEEE

Transactions on Systems, Man, and Cybernetics: Part A Systems and Humans, 2000, 30(5): 552-561.

[33] Guo XQ, Yan Qiong, Yan Qi. A Control Simulate Study of Fuzzy-neron Controller Optimized by Immune Genetic Algorithm (IGA) Used in Full Order Rectification Column Model[J]. China's New Technology and New Products, 2008, 96-99. (in Chinese)

[34] Guedria NB. Improved Accelerated PSO Algorithm for Mechanical Engineering Optimization Problems. Applied Soft Computing, 2016, 40: 455-467.

[35] Pornsing C, Sodhi MS, Lamond BF. Novel self-adaptive particle swarm optimization methods[J]. Soft Computing, 2016, 20(9): 3579-3593.

[36] Qin L, Li YQ, Zhou K. Vehicle routing problem based on heuristic artificial fish school algorithm[J]. Applied Mechanics and Materials, 2015, 3748(721): 56-61.

[37] Wang HQ, Zhang BY. Application of Fuzzy PID Controller Based on AFSA-BP Algorithm in DMF Recovery System[J]. Measurement and Control Technology, 2019, 38(10): 80-84. (in Chinese)